\documentclass[10pt,conference]{IEEEtran}
\usepackage{amsmath,amssymb}
\usepackage[alwaysadjust]{paralist}
\usepackage{cite}

\DeclareMathOperator{\wt}{wt} 
\DeclareMathOperator{\tr}{tr} 
\DeclareMathOperator{\ord}{ord} 
\DeclareMathOperator{\Gal}{Gal}

\newcommand{\F}{\mathbf{F}}
\newcommand{\Z}{\mathbf{Z}}
\newcommand{\Q}{\mathbf{Q}}

\newcommand{\qedhere}{}

\newtheorem{lemma}{Lemma} 
\newtheorem{theorem}[lemma]{Theorem} 
\newtheorem{proposition}[lemma]{Proposition} 
\newtheorem{corollary}[lemma]{Corollary}

\begin{document}
\title{Duadic Group Algebra Codes}
\author{
\authorblockN{Salah A. Aly, Andreas Klappenecker, Pradeep Kiran Sarvepalli}
\authorblockA{Department  of Computer Science, Texas A\&M University\\
College Station, TX 77843-3112, USA \\
Email: \{salah, klappi, pradeep\}@cs.tamu.edu}
}
\maketitle

\begin{abstract}
Duadic group algebra codes are a generalization of quadratic residue
codes.  This paper settles an open problem raised by Zhu concerning
the existence of duadic group algebra codes.  These codes can be used
to construct degenerate quantum stabilizer codes that have the nice
feature that many errors of small weight do not need error correction;
this fact is illustrated by an example.
\end{abstract}

\section{Introduction}
Binary cyclic duadic codes were introduced by Leon, Masley and
Pless~\cite{leon84} as a generalization of quadradic residue
codes. Duadic codes share many properties of quadratic residue codes,
but are more flexible; for example, they are not restricted to prime
lengths. Subsequently, cyclic duadic code were generalized to
nonbinary fields in~\cite{rushanan86,smid86,smid87}.  Further progress
was made by interpreting duadic codes as group algebra codes in the
sense of MacWilliams~\cite{macwilliams69}. Using this point of view,
duadic codes were extended to abelian groups by
Rushanan~\cite{rushanan86} and to nonabelian group by
Zhu~\cite{zhu96}. In this paper, we further relax the requirements and
allow for a wider class of antiautomorphisms.

An open problem by Zhu~\cite{zhu96} asks to find necessary and
sufficient conditions for the existence of duadic group algebra codes
with splitting given by $\mu_{-1}$ (the terminology is explained in the next section).
Partial answers to this question were obtained by Smid~\cite{smid87}
in the cyclic case, by Ward and Zhu~\cite{ward94} and Rushanan~\cite{rushanan86} 
in the abelian case, and by Zhang~\cite{zhang01} in the case of 
(nonabelian) finite groups and finite fields of characteristic~2.

The main result of this paper settles Zhu's question in the general
case of arbitrary finite groups and arbitrary finite fields.  We show
that if $G$ is a finite group of odd order $n$ and $\F_q$ is a finite
field with $q$ elements such that $\gcd(q,n)=1$, then there exists a
splitting given by $\mu_{-1}$ with central idempotents $e$ and $f$ if and only
if the order of $q$ is odd modulo $n$. In our proof, we establish a
key proposition that allows us to transfer the existence question of
duadic group algebra codes (for an arbitrary splitting) to a question
about so-called $\F_q$-conjugacy classes of the group. Furthermore, we
give an example for the existence of a splitting $\mu$ when
$\mu_{-1}$-splitting cannot exist. 

One of the applications of duadic
group algebra codes is the construction of (degenerate) quantum
error-correcting codes; here duadic group algebra codes can provide
even better results than cyclic duadic codes, but space constraints
will only allow us to sketch an example.
We derive a family of quantum codes with parameters $[[n,1,d]]_q$ 
such that $d^2-d+1 \geq n$.

\section{Duadic Group Algebra Codes} 
\subsection{Background}
Let $G$ be a finite group of order~$n$ with identity element $1$ 
and let $\F_q$ denote a
finite field with~$q$ elements such that $\gcd(q,n)=1$.  Recall that
the group algebra $R=\F_q[G]$ consists of elements of the form
$\sum_{g\in G} a_g g$, with $a_g\in \F_q$. The set $R$ is a vector
space over $\F_q$ in which the elements of $G$ form a basis.
Furthermore, $R$ is equipped with a multiplication defined by the
convolutional product
$$ \bigg(\sum_{g\in G} a_g g\bigg)\bigg(\sum_{g\in G} b_g g\bigg)
= \sum_{g\in G}\bigg(\sum_{h\in G}
a_{h^{\vphantom{-1}}}b_{h^{-1}g}\bigg)g. $$ A left ideal in $R$ is
an additive subgroup $I$ of $R$ such that $ra\in I$ for all $r$ in $R$
and $a$ in $I$. A \textit{group algebra code} in $R$, or shortly an
$R$-code, is a left ideal $I$ of~$R$.

For example, if $G=\Z/n\Z$ is a cyclic group, then $\F_q[\Z/n\Z]\cong
\F_q[x]/\langle x^n-1\rangle$; thus, in this case a group algebra code
is simply a cyclic code. 

An element $e$ in the group algebra $R$ is called an
\textit{idempotent} if and only if $e^2=e$. An idempotent in the
center of $R$ is called \textit{central}.  Since $\gcd(n,q)=1$, 
any $R$-code is generated by an idempotent, that is, for any
left ideal $I$ there exists an idempotent element $e$ in $R$ such that
$I=Re$. Two idempotent elements $e$ and $f$ are called
\textit{orthogonal} if $ef=0=fe$. An nonzero idempotent $e$ in $R$ is
called (\textit{centrally}) \textit{primitive} if and only if it
cannot be written as the sum of two nonzero orthogonal (central)
idempotents in $R$.

The elements $0$ and $1$ are idempotents of $R$.  If $N$ is a subgroup
of $G$, then $\widehat{N}=|N|^{-1}\sum_{g\in N} g$ is an idempotent
element in the group algebra~$R$. If $N$ is a normal subgroup, then
$\widehat{N}$ is a central idempotent.  The central idempotent
$\widehat{G}$, known as the trivial idempotent, will play a
significant role in the subsequent sections. If we multiply an element
$b=\sum_{g\in G} b_g g$ by $\widehat{G}$, then we obtain
$b\,\widehat{G}=\big(\sum_{g\in G} b_g\big) \widehat{G}$; in
particular, we have $\dim R \widehat{G}=1$. An element
$b=\big(\sum_{g\in G} b_g g\big)$ in $R$ is called \textit{even-like}
if and only if $b\,\widehat{G}=0$ (viz.  $\sum_{g\in G} b_g =0$). An
element of $R$ that is not even-like is called \textit{odd-like}.

An \textit{antiautomorphism} on the group algebra $R$ is a bijective
map $\mu$ on $R$ that satisfies (i) $\mu(a)+\mu(b)=\mu(a+b)$ and (ii)
$\mu(ab)=\mu(b)\mu(a)$ for all $a, b$ in $R$.  We say that the
antiautomorphism $\mu$ is isometric if it preserves the Hamming
weight. 
An important isometric antiautomorphism on $\F_q[G]$ is 
$\mu_{-1}$ defined as $\mu_{-1}(g)=g^{-1}$ for $g$ in $G$.

\subsection{Duadic Group Algebra Codes}
Let $G$ be a finite group of order $n$ and $\F_q$ a finite field such
that $\gcd(q,n)=1$.  If $e$ and $f$ are two even-like
idempotents in $R=\F_q[G]$ that satisfy the equations
\begin{compactenum}
\item[\textbf{A1}]\label{eq1} $e+f = 1-\widehat{G}$ and 
\item[\textbf{A2}]\label{eq2} $\mu(e)=f$ and $\mu(f)=e$ for some isometric 
antiautomorphism $\mu$ on $R$, 
\end{compactenum}
then the idempotents $e$ and
$f$ generate 
\begin{compactenum}
\item[\textbf{C1}] a pair of \textit{even-like duadic codes}
$C_e := Re$ and $C_f := Rf$, 
\item[\textbf{C2}] and a pair of \textit{odd-like duadic codes}
$D_e := R(1-f)$ and $D_f := R(1-e).$
\end{compactenum}
The antiautomorphism $\mu$ given in \textbf{A2} 
is said to give a 
splitting. By a slight abuse we also refer to $\mu_{-1}$  as a splitting.

\begin{lemma} \label{basic} 
Let $G$ be a group of order $n$ and $\F_q$ a finite field such that
$\gcd(n,q)=1$. If $e$ and $f$ are even-like idempotents in $\F_q[G]$
that satisfy \textbf{A1} and \textbf{A2} with splitting
$\mu$, then we note that
\begin{compactenum}[i)]
\item \label{basic1} 
the idempotents $e$, $f$, and $\widehat{G}$ are pairwise orthogonal;
\item \label{basic3} 
$\dim C_e=\dim C_f=(n-1)/2$ and $\dim D_e =
\dim D_f = (n+1)/2$;
\item \label{basic4} 
in particular, the order $n$ of the group $G$ must be odd;
\item \label{basic5} 
the codes satisfy the inclusions 
$C_e\subseteq D_e$ and $C_f\subseteq D_f$.
\end{compactenum}
\end{lemma}
\begin{proof} 
\ref{basic1}) Since $e$ and $f=(1-\widehat{G}-e)$ are even-like, it
follows that the idempotents $e, f$, and $\widehat{G}$ are pairwise
orthogonal; hence, $R=Re\oplus Rf\oplus R\widehat{G}$.  For \ref{basic3}) and \ref{basic4}), we observe that
$\dim R \widehat{G}=1$ and $\dim Re=\dim R\mu(e)=\dim R f$, which
implies $\dim R e = \dim Rf = (n-1)/2$.  The dimensions for the
odd-like duadic codes are an immediate consequence, since $C_e\oplus
D_f=R$ and $C_f\oplus D_e=R$. For \ref{basic5}), notice that the orthogonality
of $e$ and $f$ yields $e(1-f)=e$ and $f(1-e)=f$.  Therefore,
$C_e=Re\subset R(1-f)=D_e$ and $C_f=Rf\subset R(1-e)=D_f$.
\end{proof}

\begin{lemma}\label{isometricauto} 
Let $\F_q$ be a finite field of characteristic $p$. Suppose that $\mu$
is an isometric antiautomorphism of a group algebra $\F_q[G]$ that
satisfies $\mu(\widehat{G})=\widehat{G}$. Then there exists a Galois
automorphism $\sigma\in \Gal(\F_q/\F_p)$ and an antiautomorphism
$\mu_*$ of the group $G$ such that
\begin{equation}\label{eq:autoform} 
\mu(\alpha g) = \sigma(\alpha)\mu_*(g)\quad\text{ holds for all } 
\alpha \in \F_q,
g\in G.
\end{equation}
\end{lemma}
\smallskip
\begin{proof}
The isometry of the antiautomorphism $\mu$ implies that the image
$\mu(g)$ of an element $g$ in $G$ is of the form $\mu(g)=c_g g'$ for
some nonzero constant $c_g\in \F_q$ and $g'\in G$.  Since
$\mu(\widehat{G})=\widehat{G}$, we have
$c_g=1$ for all $g\in G$. Therefore, $\mu$ restricts to a 
antiautomorphism $\mu_*$ on the group $G$. 
Since 
$\mu$
preserves addition and multiplication of scalars and $\mu(\F_q 1)=\F_q 1$, 
we have $\mu(\alpha
1)=\sigma(\alpha) 1$ for some automorphism of $\F_q$. The elements
of the prime field $\F_p$ remain fixed, so $\sigma$ is an element of
$\Gal(\F_q/\F_p)$. The claim is an immediate consequence of these observations. 
\end{proof}

\subsection{Odd-like Weights in Duadic Group Algebra Codes}
Our first result is a slight generalization of a theorem by
Zhu~\cite{zhu96}. We allow isometric antiautomorphisms of $\F_q[G]$,
whereas Zhu considers only antiautomorphisms that are induced
from the group $G$.

\begin{lemma}\label{th:zhu}
Suppose that $e$ and $f$ are idempotents that determine a pair of
even-like duadic codes in $\F_q[G]$ with splitting given by $\mu$. 
If the group $G$ has order $n$, then the minimum weight $d_o$ of an odd-like
element in the odd-like duadic code $D_f=R(1-e)$ or $D_e=R(1-f)$ satisfies
\begin{compactenum}[i)] 
\item $d_o^2\ge n$, 
\item $d_o^2-d_o+1\ge n$ if $\mu=\mu_{-1}$. 
\end{compactenum}
\end{lemma}
\begin{proof}
i) Suppose that $a$ is an odd-like element in $D_f$ of weight $d_o$,
so there exists an element $b\in \F_q[G]$ such that $a=b(1-e)$. The
element $b$ is odd-like, since $0\neq a\widehat{G}=
b(1-e)\widehat{G}=b\widehat{G}$ holds. A splitting satisfies
$\mu(\widehat{G})=\widehat{G}$; thus, $\mu(b)$ is odd-like as well, as
$0\neq b\widehat{G}$ implies $0\neq
\widehat{G}\mu(b)=\mu(b)\widehat{G}$.

The product of $a$ and $\mu(a)$ has Hamming weight $\le
d_o^2$. However, we recall that $(1-e)(1-f)=\widehat{G}$, so
$$ a\mu(a)=b(1-e)(1-f)\mu(b)=b\widehat{G}\mu(b)=b\mu(b)\widehat{G}\neq
0.$$ 
If $b\mu(b)=\sum_{g\in G} c_g g$, then $b\mu(b)\widehat{G}=
\big(\sum_{g\in G} c_g\big) \widehat{G}$; thus, the product $a\mu(a)$
yields an element of Hamming weight $n$, which proves the bound $n\le
d_o^2$. 

For part ii), we note that $a\mu_{-1}(a)$ 
has at most Hamming weight $d_o^2-d_o+1$ when $a$ has Hamming weight~$d_o$.
By symmetry similar results hold for $D_e$.
\end{proof}

\subsection{Duals of Duadic Codes}
We can define a \textit{Euclidean inner product} on $\F_q[G]$ by
$$ 
\bigg\langle\sum_{g\in G} a_g g\bigg| 
\sum_{g\in G} b_g g\bigg\rangle
=\sum_{g\in G} a_gb_g. 
$$ 
\begin{lemma}\label{euclorth} 
Let $G$ be a finite group and $R=\F_q[G]$. 
\begin{compactenum}[i)]
\item \label{dualcrit} The product $ab=0$ for $a,b\in R$ if and 
only if $\mu_{-1}(b)\in C^\perp$, where $C=Ra$. 
\item \label{leftideal} If $C$ is an R-code, then $C^\perp$ is also an R-code.
\item \label{dualidem} If $e$ is an idempotent in $R$ and $C=Re$, then 
$C^\perp=R(1-\mu_{-1}(e))$.
\end{compactenum}
\end{lemma}
\begin{proof}
\begin{inparaenum}[i)] 
\item We note that the product of $a$ and $b$ can be expressed in the
form
$$ ab = \sum_{g\in G} \left(\sum_{h\in G} a_{gh}b_{h^{-1}} \right)g =
\sum_{g\in G} \langle g^{-1}a\mid \mu_{-1}(b)\rangle g, $$
from which we can directly deduce the claim. 

\item We note that the inner product satisfies 
$\langle ga|gb\rangle=\langle a|b\rangle$ for all
$g$ in $G$ and $a,b$ in $R$. If $a\in C$ and $b\in C^\perp$, then 
for each $g\in G$, we have 
$ \langle a | gb\rangle =\langle g^{-1}a|b\rangle=0,$ since $g^{-1}a
\in C$. Extending linearly shows that $C^\perp$ is a left ideal. 

\item Since $e(1-e)=0$, property~\ref{dualcrit}) shows
that the idempotent $1-\mu_{-1}(e)$ is contained in $C^\perp$, so
$R(1-\mu_{-1}(e))\subseteq C^\perp$ by~\ref{leftideal}).  Since
$\dim C^\perp = \dim R(1-e)=\dim R(1-\mu_{-1}(e))$, we actually must
have equality. \qedhere
\end{inparaenum}
\end{proof}

\begin{corollary} \label{co:eDuals}
If $e$ and $f$ are even-like idempotents that satisfy \textbf{A1} and \textbf{A2},
then the following statements hold:    
\begin{compactenum}[i)] 
\item If $\mu_{-1}(e)=f$, then $C_e^\perp=D_e$ and $C_f^\perp=D_f$;
\item if $\mu_{-1}(e)=e$, then $C_e^\perp = D_f$ and $C_f^\perp=D_e$. 
\end{compactenum}
\end{corollary}

\section{Existence of Splittings} \label{sec:splittings}
The goal of this section is to prove the following theorem of the
existence of duadic group algebra codes.

\begin{theorem}\label{th:existence}  
Let $G$ be a finite group of odd order and let $\F_q$ be a finite
field with $\gcd(n,q)=1$.  There exists a splitting $\mu=\mu_{-1}$
with central idempotents $e$ and $f$ such that equations \textbf{A1} 
and \textbf{A2} 
are satisfied if and only if the order of $q$ is odd
modulo $n$.
\end{theorem}
\begin{proof}(Outline) Our proof is subdivided into three parts. 
\begin{compactenum}
\item[\bf Part A.] We show that a splitting $\mu_{-1}$ with central
idempotents $e$ and $f$ satisfying \textbf{A1} and \textbf{A2} exists
if and only if no nontrivial centrally primitive idempotent is fixed
by $\mu_{-1}$, see~Proposition~\ref{pexistence}.
\item[\bf Part B.] We then define an action of $\mu_{-1}$ on so-called
$\F_{q}$-conjugacy classes in $G$. We prove that no nontrival
centrally primitive idempotent is fixed by $\mu_{-1}$ if and only if 
no $\F_q$-conjugacy class is fixed by the action of $\mu_{-1}$, see 
 Proposition~\ref{samecountNV}. 
\item[\bf Part C.]  By \cite[Lemma~2.3]{zhang01}, $K\neq \mu_{-1}(K)$
for all nontrivial $\F_q$-conjugacy classes $K$ if and only the order
of~$q$ is odd modulo~$n$.
\end{compactenum}
The claim follows by combining the three parts. 
\end{proof}

Actually, our proofs of \textbf{Part A} and \textbf{Part B} are valid
for arbitrary splittings~$\mu$. If one can find necessary and
sufficient conditions such that $\mu(K)\neq K$ holds for all nontrival
$\F_q$-conjugacy classes, then one already obtains an extension of the
theorem to $\mu$-splittings.

One technical difficulty in our proof is that the counting argument
used in \textbf{Part~B} cannot be done in nonzero characteristic. We
circumvent this problem by using an extension of the field of $p$-adic
integers (a local field) such that the ring of integers in this field
reduces to the given finite field modulo its maximal ideal; this proof
technique is interesting in itself.

\textbf{Part A.} We now supply the details of
\textbf{Part~A} of the proof. 

\begin{proposition}\label{pexistence}
There exists a splitting $\mu$ with even-like (central) idempotents $e$
and~$f$ that satisfy \textbf{A1} and \textbf{A2} if and only if each
nontrivial (centrally) primitive idempotent $h$ of $\F_q[G]$
satisfies $\mu(h)\neq h$.
\end{proposition}
\begin{proof}
Suppose that $e$ and $f$ are even-like (central) idempotents that
satisfy \textbf{A1} and \textbf{A2}.  These equations imply that
$e+f+\widehat{G}=1$, where the idempotents $e$, $f$, and $\widehat{G}$
are pairwise orthogonal. Suppose that $e=h_1+\cdots+h_m$ is a
decomposition of the idempotent $e$ into orthogonal (centrally)
primitive idempotents. Seeking a contradiction, we assume that
$\mu(h_k)=h_k$ for some $k$ in the range $1\le k\le m$. However, then
$e$ and $f$ cannot be orthogonal idempotents, contradiction.

Conversely, suppose that $h\neq \mu(h)$ for all nontrivial primitive
(central) idempotents $h$ of $\F_q[G]$.  Partition the nontrivial
(central) primitive idempotents into disjoint pairs $\{ h_1, \mu(h_1)\},
\dots, \{ h_{\ell}, \mu(h_{\ell})\}$.  Let $e=h_1+\cdots + h_\ell$ and
$f=\mu(e)$. Then $e+f=1-\widehat{G}$ and $e\widehat{G}=0$ and
$f\widehat{G}=0$. Further, $\mu(e+f+\widehat{G})=1=e+f+\widehat{G}$
implies that $e=\mu(f)$. So $\mu$ is the desired splitting with
(central) even-like idempotents $e$ and $f$.
\end{proof}

\textbf{Part B.} The second part of our argument is more involved. Our
goal is to prove the key proposition below. However, we need some
preparation to state this result. We note that a centrally primitive
idempotent $e_\chi$ of $\F_q[G]$ can be explicitly written in the form
\begin{equation}\label{eq:explidem}
 e_\chi = \frac{n_\chi}{|G|} \sum_{g\in G} \chi(g)g^{-1},
\end{equation} 
where $\chi$ is an irreducible $\F_q$-character and $n_\chi$ is a
positive integer that depends on $\chi$, see Lemma~\ref{th:idemform}
in the Appendix. 
\begin{lemma}\label{th:muAction}
Let $\F_q$ be a finite field and $G$ a finite group of order $n$ such
that $\gcd(n,q)=1$.  Suppose that $\mu$ is an antiautomorphism of
$\F_q[G]$ of the form (\ref{eq:autoform}). Then the action of $\mu$ on
a centrally primitive idempotent (\ref{eq:explidem}) is given by
$$ \mu(e_\chi) = \frac{n_\chi}{|G|}\sum_{g\in G} \chi(\mu_*^{-1}(g^k))g^{-1},
$$ where $k$ is a positive integer determined by the Galois
automorphism $\sigma$ and $\mu_*^{-1}$ is the inverse of the group
antiautomorphism $\mu_*$. In particular, $k$ is a power of the
characteristic of\/ $\F_q$.
\end{lemma}
\begin{proof}
If the exponent of the group $G$ is $m$, then the values of the
character are contained in $\F_q \cap \F_p(\delta)$, where $\delta$ is
a primitive $m$-th root of unity over the prime field $\F_p$. Suppose
that $\sigma'$ is a Galois automorphism of $\Gal(\F_q(\delta)/\F_p)$
that restricts to the Galois automorphism $\sigma$ on $\F_q$. If
$\sigma'(\delta)=\delta^k$, then $\sigma(\chi(g))=\chi(g^k)$ holds for
all $g$ in $G$, and the action of an antiautomorphism $\mu$ is given
by
$$ \mu(e_\chi) = \frac{n_\chi}{|G|}\sum_{g\in G} \sigma(\chi(g))
\mu_*(g^{-1}) = \frac{n_\chi}{|G|}\sum_{g\in G} \chi(\mu_*^{-1}(g^k))
g^{-1},
$$ where the latter equality is obtained by substituting
$\mu_*^{-1}(g)$ for $g$. 
\end{proof}

Let us recall the concept of an $\F_q$-conjugacy class before stating
our next result. The $\F_q$-conjugacy class $K_q(g)$ of an element $g$
in a finite group $G$ is the set
$$ K_q(g) = \{ h^{-1}g^{q^k}h\,|\, h \in G, k\ge 0\}.$$ It is easy to
see that two $\F_q$-conjugacy classes are either disjoint or coincide.
The following two key facts are essential for our purpose: 
\begin{inparaenum}[(i)]\item An
irreducible $\F_q$-character is constant on $\F_q$-conjugacy classes,
and \item the number of $\F_q$-conjugacy coincides with the number of
irreducible $\F_q$-characters.
\end{inparaenum}

We define an action of an antiautomorphism $\mu$ of the form
(\ref{eq:autoform}) on an $\F_q$-conjugacy class $K_q(g)$ by
$$ K_q^\mu(g) = K_q(\mu_*(g)^{\ell}),$$ where $\ell$ is a positive
integer such that $k\ell\equiv 1\bmod m$, $k$ is the integer given in
Lemma~\ref{th:muAction}, and $m$ is the exponent of the group $G$.

\begin{proposition}[Key Proposition]\label{samecountNV} 
Let $\F_q$ be a finite field and let $G$ be a 
finite group of order $n$ such that $\gcd(n,q)=1$.  If $\mu$ is an
antiautomorphism of $\F_q[G]$ of the form 
(\ref{eq:autoform}), then the number of\/ $\F_q$-conjugacy classes of\/ 
$G$ that are fixed by $\mu$ coincides with the number of
centrally primitive idempotents of $\F_q[G]$ that are fixed by $\mu$. 
\end{proposition}
\begin{proof} 
\begin{inparaenum}
\item[\it Step 1.] Suppose that the finite field $\F_q$ has
characteristic $p$.  There exists a monic irreducible polynomial
$f(x)$ in $\F_p[x]$ such that $\F_p[x]/\langle f(x)\rangle = \F_q$.
Let $\Z_p$ denote the ring of $p$-adic integers and $\Q_p$ its
quotient field. Then $\mathfrak{p}=p\Z_p$ is the unique nonzero prime
ideal of $\Z_p$ and $\Z_p/\mathfrak{p}\cong \F_p$.  Choose a monic
polynomial $g(x)$ in $\Z_p[x]$ such that $f(x)\equiv g(x)\mod
\mathfrak{p}$.  Then $R=\Z_p[x]/\langle g(x)\rangle$ is a discrete
valuation ring with nonzero prime ideal $\mathfrak{P}=\mathfrak{p}R$.

If $K$ denotes the field $\Q_p[x]/\langle g(x)\rangle$, then $K/\Q_p$
is an unramified Galois extension. The Galois group $\Gal(K/\Q_p)$ can
be identified with the cyclic group $\Gal(\F_q/\F_p)$. The latter
group is generated by the automorphism $x\mapsto x^p$, and the
generator of $\Gal(K/\Q_p)$ can be characterized by
$$ b\equiv b^p \bmod \mathfrak{P}\quad \text{for all } \quad b \in R.$$ 
All facts stated in this step are proved in \cite[Chapter 1]{serreLF}.

\item[\it Step 2.]  The number of centrally primitive idempotents in
$K[G]$ and in $\F_q[G]$ is the same. If $Y$ denotes the set of
irreducible $K$-characters of $G$, then the set of centrally primitive
idempotents $\{ e_\chi\,|\, \chi\in Y\}$ of $K[G]$ is bijectively
mapped to the centrally primitive idempotents of $\F_q[G]$ by
reduction modulo $\mathfrak{P}$.

Let $k$ be the integer defined as in Lemma~\ref{th:muAction}. Then
there exists a unique automorphism $\tau$ in $\Gal(K/\Q_p)$ such that
$\tau(x)=x^k\bmod \mathfrak{P}$. Therefore, we can define an
anti\-automorphism $\eta$ on $K[G]$ by $\eta(\alpha g) =
\tau(\alpha)\mu_*(g)$ such that 
\begin{equation}\label{eq:idemreduc}
\eta(e_\chi) \bmod \mathfrak{P} = \mu(e_\chi \bmod \mathfrak{P})
\end{equation}
holds for all centrally primitive idempotents $e_\chi$ of $K[G]$.  The
latter equation guarantees that the number of idempotents in $K[G]$
fixed by $\eta$ is the same as the number of idempotents in $\F_q[G]$
fixed by $\mu$.

\item[\it Step 3.]
A centrally primitive idempotent in $K[G]$ is of the form 
$$ e_\chi = \frac{n_\chi}{|G|} \sum_{g\in G} \chi(g)g^{-1}.$$ It
follows from Lemma~\ref{th:muAction} that $\eta(e_\chi)=e_\chi$ if
and only if $\chi(g)=\chi(\eta^{-1}(g^k))$ holds for all $g$ in $G$.
Therefore, we define the action of $\eta$ on an irreducible
$K$-character by
\begin{eqnarray}
\chi^{\eta}(g) &=& \chi(\mu_*^{-1}(g^k)), \label{eq:charAction} 
\end{eqnarray}
for all $g$ in $G$. 

An irreducible $K$-character is constant on $K$-conjugacy classes. The
$K$-conjugacy classes coincide with the $\F_q$-conjugacy classes,
since the Galois groups are isomorphic.  Suppose that $m$ is the
exponent of the group $G$.  There exists a positive integer $\ell$
such that $k\ell\equiv 1\bmod m$.  We define the action of $\eta$ on
$\F_q$-conjugacy classes by
\begin{eqnarray}
K_q^\eta(g) &=& K_q(\mu_*(g^\ell)), \label{eq:conjAction} 
\end{eqnarray}
for all $g$ in $G$. The definitions are carefully chosen such that
$$ \chi^\eta(K_q^\eta(g)) = \chi(K_q(g))$$
holds for all $g$ in $G$.  

\item[\it Step 4.] Let $K_q$ denote the set of $\F_q$-conjugacy classes. 
We have $|Y|=|K_q|$. Therefore, we can define the square matrix
$$ 
U = \left(\chi(K)\right)_{\chi\in Y,K\in K_q}. 
$$ 
We note that $U$ is nonsingular, since the irreducible $K$-characters
are linearly independent over $K$.  Let
$A=(A_{\chi,\psi})_{\chi,\psi\in Y}$ and $B=(B_{L,K})_{L,K\in K_q}$
be permutation matrices that are respectively defined by 
$$ A_{\chi,\psi}=\left\{\begin{array}{ll}
1 & \text{if } \chi=\psi^{\eta}\\
0 & \text{otherwise}
		       \end{array}
\right.
\quad \text{and} \quad
B_{L,K}=\left\{\begin{array}{ll}
1 & \text{if } K^{\eta}=L\\
0 & \text{otherwise}.
		       \end{array}
\right. 
$$ 
Since 
$\chi^{\eta}(K^{\eta})=\chi(K)$, we have
$$ \sum_{\psi\in Y} A_{\chi,\psi} \psi(K) = \chi(K^{\eta})
\quad\text{and}\quad
\sum_{L\in K_q} \chi(L) B_{L,K} = \chi(K^{\eta}), $$ so
$AU=UB$. Since $U$ is invertible, we have $A=UBU^{-1}$. Thus, 
$\tr(A)=\tr(B)$. The trace of $A$ counts the number of
characters that remain fixed under the action of $\eta$, and the 
trace of $B$ counts the number of $\F_q$-conjugacy classes that remain
fixed under $\eta$. These facts imply the claim. \qedhere
\end{inparaenum}
\end{proof}

Recall that \textbf{Part C} has been proved in \cite{zhang01}; thus,
this concludes our proof of Theorem~\ref{th:existence}.  For the
existence of duadic group algebra codes with splitting $\mu\neq
\mu_{-1}$ one only needs to modify~\textbf{Part C}.

\section{Extensions and Applications}\label{sec:quantumcodes}
The natural question following the previous section is the existence
of duadic group algebra codes when the order of $q$ is even. The
splitting is no longer given by $\mu_{-1}$, but we will show that
there exist duadic algebra codes.  We will confine ourselves to the
abelian case.  Then we will give an application of duadic group
algebra codes to quantum error-correction.

\subsection{Extensions}
Cyclic duadic codes exist if and only if $q$ is a quadratic
residue modulo $n$.  However, this condition is not required for group
codes.  We partially generalize some of the existence results of
\cite{rushanan86}, where characteristic two is considered.
\begin{lemma}\label{th:mu4nonresidue}
Let $G=\langle a,b\,|\, a^p=b^p=1, ab=ba\rangle$, where $p$ is an odd
prime and $q$ be a prime power such that $\ord_p(q)$ is even and
$\gcd(p,q)=1$. The $\F_q$-conjugacy class of an element $a^xb^y$ in
$G$ is given by $C_{x,y}^{(q)}=\{ a^{xq^j}b^{yq^j} \mid j\in \Z\}$.
Then the automorphism $\mu$ defined as $\mu(a^xb^y)=a^{qy}b^x$ does
not fix any of the $\F_q$-conjugacy classes if $(x,y)\neq (0,0)$.
Further, $\mu_{-1}$ fixes each $\F_q$-conjugacy class {\em i.e.},
$\mu_{-1}(C_{x,y}^{(q)})= C_{x,y}^{(q)}$.
\end{lemma}
\begin{proof}
Assume that there is an element $a^xb^y\in G$ , such that
$\mu(C_{x,y}^{(q)})=C_{x,y}^{(q)}$.  Then there exists an integer $j$
such that $\mu(a^xb^y)=a^{xq^j}b^{yq^j}$. This implies that
$(qy,x)=(xq^j,yq^j) \mod p$ or $qy\equiv xq^j\mod p$ and $x\equiv
yq^j\mod p$. If $x=0, y\neq 0$, then we have $qy\equiv 0\mod p$ or
$y=0$; a contradiction. If $y=0,x\neq 0$, then it follows $x=0$, which
leads to a contradiction again. Assuming that both $x,y\neq 0$ we get
$qxy\equiv xyq^{2j} \mod p$. Since $q,x,y$ are all coprime to $p$ this
can be written as $1\equiv q^{2j-1}\mod p$. But as $\ord_p(q)$ is
even, $1\not\equiv q^{2j-1}\mod p$.  Therefore, none of the $\F_q$-conjugacy
classes are fixed by $\mu$.  Let $\ord_p(q)=2w$, then $q^{2w}\equiv 1
\mod p$, which implies that $q^w\equiv -1\mod p$. Hence,
$C_{x,y}^{(q)}=C_{-x,-y}^{(q)}=\mu_{-1}(C_{x,y}^{(q)})$.
\end{proof}

\begin{theorem}\label{th:qnonresidue}
Let $G=\langle a,b\,|\, a^p=b^p=1, ab=ba\rangle$ with $p$ an odd
prime, $q$ a prime power, $\gcd(p,q)=1$ and $\ord_p(q)$ even. Then
there exist duadic codes over $\F_q[G]$ with splitting given by $\mu$
where $\mu(a^xb^y)=a^{qy}b^x$ for any element $a^xb^y\in G$. These
codes are fixed by $\mu_{-1}$.
\end{theorem}
\begin{proof}
By \textbf{Part A} (Proposition~\ref{pexistence}), \textbf{Part B}
(Proposition~\ref{samecountNV}), and replacing \textbf{Part C} by
Lemma~\ref{th:mu4nonresidue}, we know that there exist a pair of
duadic codes over $\F_{q}[G]$ with splitting given by $\mu$ such that
the codes are fixed by $\mu_{-1}$.
\end{proof}

\subsection{Quantum Error-Correction}
Quantum codes can be utilized to protect quantum information over
noisy quantum channels.  The CSS construction is particularly
transparent method to derive quantum stabilizer codes from a pair of
classical codes.
\begin{lemma}[CSS Construction~\cite{calderbank96,steane96}]\label{lm:css}
Suppose that $C$ and $D$ are linear codes over a finite field~$\F_q$
such that $C\subseteq D$. If $C$ is an $[n,k_1]_q$ code and $D$ an
$[n,k_2]_q$ code, then there exists an $[[n,k_2-k_1,d]]_q$ stabilizer
code with minimum distance $d=\min \wt\{ (D\setminus C)\cup
(C^\perp\setminus D^\perp)\}.$
\end{lemma}

\begin{theorem}
Let $n$ be an odd positive integer and $q$ a power of a prime such
that the order of $q$ modulo $n$ is odd. Then there exists an
$[[n,1,d]]_q$ stabilizer code with $d^2-d+1\ge n$.
\end{theorem}
\begin{proof}
There exists a group $G$ of order $n$. By Theorem~\ref{th:existence},
there exist idempotents $e$ and $f$ in $\F_q[G]$ that satisfy
\textbf{A1} and \textbf{A2} with splitting given by $\mu=\mu_{-1}$. By
Lemma~\ref{basic}, we have $C_e\subset D_e$.  The CSS construction
shows that there exists an $[[n,(n+1)/2-(n-1)/2,d]]_q=[[n,1,d]]_q$
stabilizer code with minimum distance $d=\min\{ (D_e\setminus C_e)
\cup (C_e^\perp\setminus D_e^\perp)\} =\min\{ (D_e\setminus C_e) \cup
(D_e\setminus C_e)\}$, where the latter equality follows from
Corollary~\ref{co:eDuals}~i).  Since $D_e=R\widehat{G}\oplus C_e$, we
observe that $D_e\setminus C_e$ contains precisely the odd-like
elements of $D_e$.  By Lemma~\ref{th:zhu}, the minimum weight of an
odd-like element in $D_e$ satisfies $d^2-d+1\ge n$.
\end{proof}

The distance of the quantum codes does not depend
on $\wt(C)$ or $\wt(D)$ or even their dual distances, rather the set
differences $D\setminus C$ and $C^\perp\setminus D^\perp$.  This means
that a code that is bad in the classical sense can lead to a good
quantum code, if only we can arrange the low weight codewords of $D$
to be entirely in $C$ and similarly for the low weight codewords of
$C^\perp$ to be in $D^\perp$; this phenomenon is called degeneracy.  
A nice consequence of degeneracy is
that errors in $C$ or $D^\perp$ do not require any error-correction,
which is a desirable feature as quantum error-correction can be
faulty. Thus we require many likely errors to be in $C$ and
$D^\perp$. Of course, it is difficult to construct codes that satisfy
this strange requirement.

Duadic group algebra codes can meet these conflicting requirements,
because their odd-like distance grows with the length $n$, while we
can design their even-like distance to be very small.  In
\cite{aly06b}, we showed how this can be done for cyclic duadic codes
over $\F_q$. These codes exists if and only if $q$ is a quadratic
residue modulo $n$. Duadic group algebra codes on the other hand
enable us to overcome this restriction. 

Now we will give some evidence of the usefulness of the group algebra
codes, by constructing a degenerate $[[81,1,\geq 9]]_2$ quantum code
which has many codewords of weight $4$.  Let $G_i=\Z_3\times \Z_3
=\langle a_i,b_i\,|\, a_i^3=b_i^3=1, a_ib_i=b_ia_i\rangle$, then from Theorem~\ref{th:qnonresidue} we
know that there exist duadic group algebra codes over $\F_2[G_i]$ with
idempotents $e_i=a_i+a_i^2+a_ib_i+a_i^2b_i^2$ and
$f_i=b_i+b_i^2+a_ib_i^2+a_i^2b_i$ satisfying \textbf{A1} and
\textbf{A2}, under the action of
$\mu_i(a_i^xb_i^y)=a_i^{2y}b_i^x$. Further $e_i,f_i$ are fixed by
$\mu_{-1}$.

We can construct a duadic code over $\F_2[G_1\times G_2]$ using a
construction similar to \cite[Theorem~5.6]{rushanan86}. Embedding
$G_i$ into $G_1\times G_2$, we get the idempotents of the new code as
$e=e_1+e_2-e_1e_2-f_1e_2$ and $f=f_1+f_2-f_1f_2-e_1f_2$.  The
splitting for this code is given by $\mu =\mu_1\times\mu_2$.

These idempotents give us duadic group algebra codes over
$\F_2[G_1\times G_2]$ that are fixed by $\mu_{-1}$.  As $ee_1=e_1$,
$\wt(C_e)=\wt(C_f)\leq \wt(e_1)=4$, while $\wt(D_e\setminus C_e)=
\wt(C_e^\perp\setminus D_e^\perp) =\wt(D_f\setminus C_f)\geq 9$ by
Corollary~\ref{co:eDuals}~ii) and Lemma~\ref{th:zhu}.  Thus by
Lemma~\ref{lm:css} there exists a $[[81,1,\geq 9]]_2$ quantum code;
this code is degenerate and many errors of weights between $4$ and~$9$
(contained in $C_e$ or $D_e^\perp$) do not need error-correction.

\appendix
\begin{lemma}\label{th:idemform}
Let $G$ be a finite group of order $n$ and $F$ a finite field of
characteristic coprime to $n$. If $E\supseteq F$ is a splitting field
for $G$, $W$ an irreducible $FG$-modul affording the character
$\chi$, and $V$ an irreducible submodul of the $EG$-modul $E\otimes W$
affording the character $\theta$, then
$ \chi(x) = \sum_{\sigma\in H} \theta^\sigma(x),$
where $H=\textup{Gal}(\F_q(\theta)/\F_q)$, and 
$ e = \frac{\theta(1)}{|G|} \sum_{g\in G} \chi(g^{-1})g$
is a centrally primitive idempotent in $FG$.  
\end{lemma}
\begin{proof}
Since $F$ is a finite field, we note that the Schur index of the
character $\theta$ is 1 by~\cite[Theorem 24.10]{dornhoffI}. Therefore,
the character $\chi$ is of the claimed form by~\cite[Theorem
24.14]{dornhoffI}. We note that 
$ e_\theta = \frac{\theta(1)}{|G|}\sum_{g\in G} \theta(g^{-1})g$ is a
centrally primitive idempotent of $EG$ by \cite[Lemma
24.13]{dornhoffI}. The form of the character $\chi$ implies that
$e=\sum_{\sigma \in H} e_{\theta^\sigma}$, so $e$ is a central
idempotent of $FG$. The primitivity of $e$ is shown in
\cite[Theorem~19.2.9]{karpilovskyIB}.
\end{proof}

\scriptsize 
\def\cprime{$'$}

\end{document}